\documentclass{article}
\usepackage{spconf,amsmath,amssymb,graphicx}
\usepackage[hidelinks]{hyperref}
\usepackage{booktabs,multirow}

\newcommand{\reimpl}{\rlap{\textsuperscript{\ensuremath{\dagger}}}}
\newcommand{\arena}{\rlap{\textsuperscript{*}}}

\title{DuRe-ST: Dual-Relation Spectro-Temporal Modeling for Speech Deepfake Detection}

\name{Shaole Li, Siqing Qin, Youzhi Tu, and Kong Aik Lee$^{\dagger}$%
\thanks{$^\dagger$ Corresponding author: Kong Aik Lee.}}

\address{
Department of Electrical and Electronic Engineering\\
The Hong Kong Polytechnic University, Hong Kong SAR
}

\begin{document}
\ninept
\maketitle

\begin{abstract}
Previous speech deepfake detectors can adaptively capture spectro-temporal dependencies through graph attention, yet they largely overlook the co-variation between spectral and temporal representations. To address this gap, we construct a normalized affinity graph from their joint covariance and apply polynomial graph filtering to capture higher-order covariance-induced dependencies. We first develop Cov-ST to isolate the contribution of covariance-based relational modeling. Although it improves detection performance, its sensitivity to the polynomial order suggests limited robustness when covariance relations are modeled alone. We therefore propose DuRe-ST, which jointly exploits covariance-induced and graph-attention-induced relations to capture complementary second-order co-variation and adaptive spectro-temporal dependencies. Experiments show that DuRe-ST achieves an average relative EER reduction of 25.9\% over XLSR-AASIST on the ASVspoof benchmarks and 28.4\% across four cross-dataset benchmarks with only 4--8k additional trainable back-end parameters. It further outperforms the strongest publicly available comparison models by 2.2--13.2\% in relative EER on four benchmarks, while remaining smaller than the publicly available models considered.
\end{abstract}


\section{Introduction}
\label{sec}

Deepfake speech generated by text-to-speech (TTS) and voice conversion (VC) methods raises concerns about voice impersonation, fraud, and misinformation~\cite{liu2023asvspoof,du2024dfadd}. These risks have driven advances in speech deepfake detection (SDD), including new model architectures~\cite{jung2022aasist,truong2024tcm}, training strategies~\cite{regularization,dpda_ga}, and data augmentation methods~\cite{tak2022rawboost}. Many recent detectors use pretrained self-supervised speech models, such as XLS-R~\cite{babu2022xlsr} and WavLM~\cite{chen2022wavlm}, as front-end feature extractors and rely on a task-specific back-end to model spoofing cues.

These back-ends model spectral and temporal information in different ways. Convolutional models, such as CG-Res2Net~\cite{li2021cgres2net} and Nes2Net~\cite{liu2025nes2net}, capture local acoustic patterns through multi-scale feature extraction. Transformer- and Conformer-based models use self-attention to model long-range context~\cite{truong2024tcm,rosello2023conformer}, while Mamba-based detectors employ selective state-space models for efficient sequence modeling~\cite{xiao2025xlsrmamba}. AASIST takes a different approach by organizing spectral and temporal features as heterogeneous graph nodes and using graph attention to model
relations within and across the two domains~\cite{jung2022aasist,tak2022wav2vec}. As a mainstream backbone, this heterogeneous modeling makes AASIST natural for studying how spectral and temporal features interact in SDD.

However, AASIST models spectro-temporal relations mainly through learned attention weights. These weights capture the relevance between nodes but do not explicitly represent covariance, a second-order statistic that describes how spectral and temporal features vary together. This omission matters because synthetic speech may contain inconsistent or unnaturally over-smoothed time--frequency structures~\cite{gao2021generalized,liu2022spectrotemporal,ren2022oversmooth}. This suggests that speech generation models do not always preserve natural spectro-temporal correlations, and deviations from these correlations may manifest as artifact traces. Such artifacts may therefore provide spoofing cues complementary to attention-induced relations, yet remain underexplored in existing SDD models. Covariance neural networks (VNNs) provide a learnable approach to exploiting
covariance information through polynomial filtering~\cite{sihag2022covariance}.

To address this gap, we construct a normalized affinity graph from the joint covariance of spectral and temporal representations and apply polynomial graph filtering to capture multi-hop covariance-induced dependencies. Based on this formulation, we first develop Cov-ST to examine the contribution of covariance-induced relation modeling. Although Cov-ST improves detection performance, it is sensitive to the polynomial order, indicating limited robustness when covariance relations are modeled in isolation. We therefore propose DuRe-ST, which jointly exploits covariance-induced and graph-attention-induced relations in a complementary manner. This dual-relation modeling captures both adaptive dependencies between spectral and temporal nodes and co-variation among their high-level representations. The main contributions of this work are as follows:
\begin{itemize}
\item We introduce covariance-induced spectro-temporal relation modeling to explicitly capture second-order dependencies between spectral and temporal representations.
\item We propose Cov-ST for covariance-only relation modeling and DuRe-ST for complementary fusion with graph-attention relations.
\item Experiments demonstrate that covariance-induced relations capture
discriminative spoofing cues complementary to attention-induced relations, enabling DuRe-ST to achieve the best results among the compared systems
across multiple benchmarks.
\end{itemize}

\section{Proposed DuRe-ST Architecture}
\label{sec:method}

\begin{figure*}[t]
    \centering
    \includegraphics[width=1\textwidth]{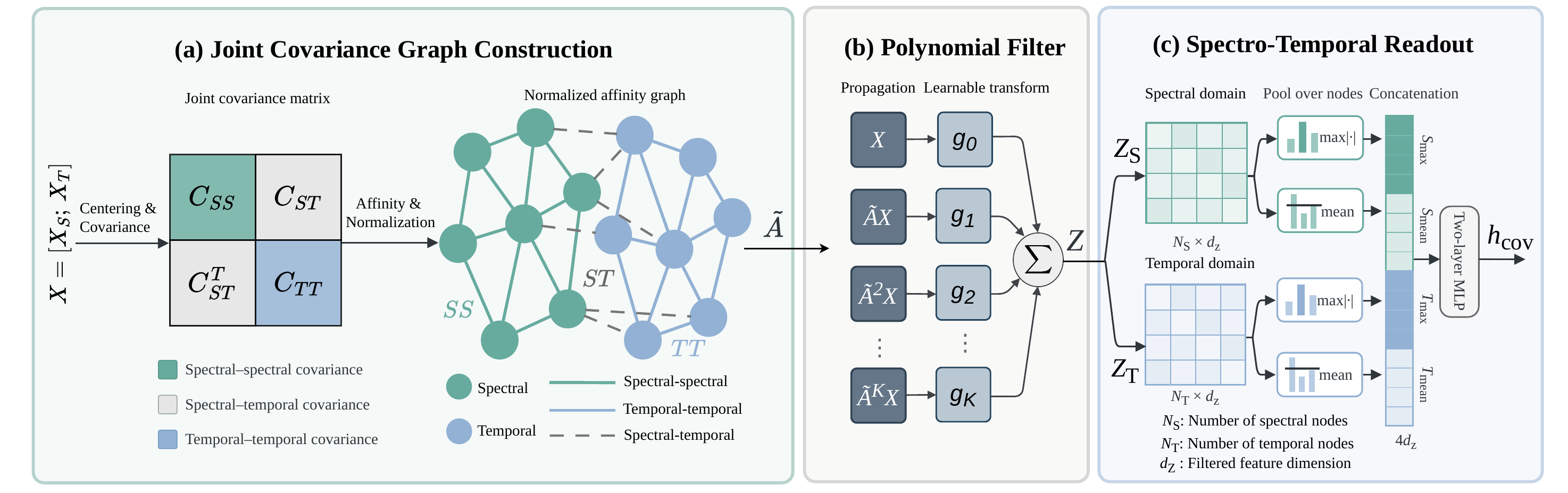}
    \vspace{-10pt}
    \caption{Overview of covariance-induced spectro-temporal relation modeling.
    Given the shared node features
    $\mathbf{X}=[\mathbf{X}_{\mathrm{S}};\mathbf{X}_{\mathrm{T}}]$,
    (a) constructs a normalized affinity graph $\widetilde{\mathbf{A}}$
    from their joint covariance, capturing second-order relations within and
    across the spectral and temporal domains;
    (b) applies an order-$K$ polynomial graph filter over
    $\widetilde{\mathbf{A}}$ to aggregate direct and multi-hop
    covariance-induced dependencies, yielding $\mathbf{Z}$; and
    (c) partitions $\mathbf{Z}$ into the spectral and temporal node
    representations $\mathbf{Z}_{\mathrm{S}}$ and $\mathbf{Z}_{\mathrm{T}}$,
    and independently summarizes each domain by max-magnitude and mean pooling
    over nodes. The four pooled statistics are concatenated and projected by a
    two-layer MLP to obtain $\mathbf{h}_{\mathrm{cov}}$.
    Cov-ST uses $\mathbf{h}_{\mathrm{cov}}$ for classification, whereas
    DuRe-ST concatenates it with the attention-induced representation
    $\mathbf{h}_{\mathrm{att}}$ to form
    $\mathbf{h}_{\mathrm{dual}}$ for classification.}
    \label{fig:dure_st}
    \vspace{-8pt}
\end{figure*}

\subsection{Shared Spectro-Temporal Nodes}

Following~\cite{tak2022wav2vec}, we use pretrained XLS-R
300M~\cite{babu2022xlsr} followed by a linear projection and a residual
CNN encoder to obtain $\mathbf{H}\in\mathbb{R}^{C\times F\times T}$,
where $C$, $F$, and $T$ denote the channel, frequency, and temporal dimensions. A 2-D self-attentive module produces logits, normalized by softmax along
the temporal and frequency axes to yield $\mathbf{W}_{\mathrm{S}}$
and $\mathbf{W}_{\mathrm{T}}$. The corresponding representations are
\begin{equation}
\mathbf{U}_{\mathrm{S}}
=\Bigl(\sum_t \mathbf{W}_{\mathrm{S}}\odot\mathbf{H}\Bigr)^{\top},
\quad
\mathbf{U}_{\mathrm{T}}
=\Bigl(\sum_f \mathbf{W}_{\mathrm{T}}\odot\mathbf{H}\Bigr)^{\top},
\label{eq:shared_nodes}
\end{equation}
where $\odot$ denotes element-wise multiplication and the sums operate
along the temporal and frequency axes, respectively.

We omit AASIST's separate spectral and temporal GAT modules and directly
apply top-$k$ graph pooling to $\mathbf{U}_{\mathrm{S}}$ and
$\mathbf{U}_{\mathrm{T}}$, avoiding prior attention-specific transformations.
Pooling reweights nodes by learned scores and retains the highest-scoring
nodes, yielding compact representations
$\mathbf{X}_{\mathrm{S}}\in\mathbb{R}^{N_{\mathrm{S}}\times C}$ and
$\mathbf{X}_{\mathrm{T}}\in\mathbb{R}^{N_{\mathrm{T}}\times C}$.
Their concatenation
$\mathbf{X}=[\mathbf{X}_{\mathrm{S}};\mathbf{X}_{\mathrm{T}}]$
is shared by both relation branches.

\subsection{Attention-Induced Relation Modeling}

For attention-induced relation modeling, we adopt the heterogeneous
graph-attention module of AASIST~\cite{jung2022aasist,tak2022wav2vec},
which comprises a master node and two successive heterogeneous
graph-attention layers. The module takes the shared spectral and temporal
representations $\mathbf{X}_{\mathrm{S}}$ and $\mathbf{X}_{\mathrm{T}}$
as input and models intra-spectral, intra-temporal, and
cross-spectro-temporal relations. Finally, the spectral and temporal
representations are summarized by global maximum and average pooling and
concatenated with the final master-node representation to form
$\mathbf{h}_{\mathrm{att}}$.

\subsection{Covariance-Induced Relation Modeling}

The covariance-induced modeling comprises joint covariance graph construction,
polynomial graph filtering, and spectro-temporal readout, as illustrated in
Fig.~\ref{fig:dure_st}(a)--(c).

We treat the shared spectro-temporal nodes in $\mathbf{X}$ as variables
and their $d = C$ feature dimensions as observations. After centering each node
across its feature dimensions, we compute the joint covariance matrix:
\begin{equation}
\mathbf{C}
=
\frac{1}{d-1}\bar{\mathbf{X}}\bar{\mathbf{X}}^{\top}
=
\begin{bmatrix}
\mathbf{C}_{\mathrm{SS}} & \mathbf{C}_{\mathrm{ST}}\\
\mathbf{C}_{\mathrm{ST}}^{\top} & \mathbf{C}_{\mathrm{TT}}
\end{bmatrix},
\label{eq:joint_covariance}
\end{equation}
where $\bar{\mathbf{X}}$ denotes the row-wise centered representation.
The diagonal blocks $\mathbf{C}_{\mathrm{SS}}$ and
$\mathbf{C}_{\mathrm{TT}}$ capture within-domain covariance, while
$\mathbf{C}_{\mathrm{ST}}$ captures cross-domain co-variation.

To model dependency strength irrespective of sign, we use covariance
magnitudes as graph affinities, add self-connections, and apply symmetric
degree normalization:
\begin{equation}
\mathbf{A}=\operatorname{abs}(\mathbf{C})+\mathbf{I}_{N},
\qquad
\widetilde{\mathbf{A}}
=
\mathbf{D}_{A}^{-\frac{1}{2}}\mathbf{A}\mathbf{D}_{A}^{-\frac{1}{2}}.
\label{eq:normalized_covariance_graph}
\end{equation}
The element-wise absolute value retains the strength of both positive and
negative co-variation and avoids cancellation caused by oppositely signed
edge weights during aggregation. Here, $\mathbf{I}_{N}$ is the identity
matrix, $N$ is the total number of spectral and temporal nodes, and
$\mathbf{D}_{A}$ is the degree matrix whose diagonal entries are the row
sums of $\mathbf{A}$.

An order-$K$ polynomial graph filter then transforms $\mathbf{X}$ over
the normalized graph $\widetilde{\mathbf{A}}$:
\begin{equation}
\mathbf{Z}
=
\operatorname{SELU}\!\left(
\operatorname{LN}\!\left(
\sum_{k=0}^{K}
g_k\!\left(\widetilde{\mathbf{A}}^{k}\mathbf{X}\right)
\right)
\right),
\label{eq:polynomial_filter}
\end{equation}
where $g_k(\cdot)$ denotes an order-specific linear transformation and
$\operatorname{LN}$ denotes layer normalization. The zeroth-order term
transforms each node independently, while terms with $k\geq1$ aggregate
features through covariance-induced relations over up to $K$ propagation
steps.

Finally, we split $\mathbf{Z}$ into spectral and temporal representations
and summarize each domain using maximum-magnitude and mean pooling across
nodes. For $q\in\{\mathrm{S},\mathrm{T}\}$,
\begin{equation}
\mathbf{p}_{q}
=
\operatorname{Max}\!\left(\operatorname{abs}(\mathbf{Z}_{q})\right)
\mathbin{\Vert}
\operatorname{Mean}\!\left(\mathbf{Z}_{q}\right),
\label{eq:covariance_pooling}
\end{equation}
where $\Vert$ denotes feature concatenation. The two summaries are
concatenated and projected to obtain the covariance-induced representation:
\begin{equation}
\mathbf{h}_{\mathrm{cov}}
=
f_{\mathrm{proj}}\!\left(
\mathbf{p}_{\mathrm{S}}\mathbin{\Vert}\mathbf{p}_{\mathrm{T}}
\right),
\label{eq:covariance_representation}
\end{equation}
where $f_{\mathrm{proj}}$ is a two-layer MLP with ReLU activations and dropout.

\subsection{Model Variants and Classification}

Building on the relation representations derived above, we propose two
models. Cov-ST uses only the covariance-induced representation, whereas
DuRe-ST jointly exploits the covariance- and attention-induced
representations. For DuRe-ST, $\mathbf{h}_{\mathrm{cov}}$ and
$\mathbf{h}_{\mathrm{att}}$ are concatenated to form the dual-relation
representation $\mathbf{h}_{\mathrm{dual}}$. The corresponding representation
of each model is subsequently mapped by a linear classifier to the bona fide
and spoof logits.

\section{Experimental Setup}
\label{sec:experiments}
\subsection{Datasets and Metric}

All models developed in this work are trained exclusively on the ASVspoof 2019 LA training
partition, which contains bona fide speech and spoofed utterances generated
by six TTS/VC algorithms~\cite{wang2020asvspoof2019}. In this work, ASVspoof 2021 LA and DF~\cite{liu2023asvspoof} are used for the ASVspoof benchmark evaluation, whereas In-the-Wild
~\cite{muller2022wild}, the normalized Fake-or-Real test
set\footnote{\url{https://www.kaggle.com/datasets/mohammedabdeldayem/the-fake-or-real-dataset}}~\cite{reimao2019for},
ASVspoof~5~\cite{wang2024asvspoof5}, and DFADD~\cite{du2024dfadd} are used
for the cross-dataset evaluation. All evaluations are conducted
without target-domain adaptation. Performance is reported using equal error
rate (EER, \%), where lower values indicate better detection performance.
\vspace{-5pt}
\subsection{Implementation Details}
\vspace{-3pt}
During training, each utterance is cropped or repeatedly concatenated to approximately 4 seconds (64,600 samples). To mitigate the class imbalance between bona fide and spoofed samples, we employ weighted cross-entropy loss with class weights of 0.9 and 0.1 for bona fide and spoofed speech, respectively~\cite{jung2022aasist,xiao2025xlsrmamba}. Models are optimized using Adam with a learning rate of $10^{-6}$, a weight decay of $10^{-5}$, and a batch size of 16. Training stops after seven epochs without improvement in validation
cross-entropy; we average the five checkpoints with the lowest validation
losses for evaluation. All experiments are conducted on a single RTX~4090 GPU.

Following prior work~\cite{tak2022rawboost,tak2022wav2vec,xiao2025xlsrmamba}, we train separate checkpoints using RawBoost Algo~3 and Algo~5. Algo~3 is used for ASVspoof 2021 DF and Algo~5 for the other datasets, while both are evaluated on DFADD and the lower EER is reported.
\vspace{-5pt}
\section{Results and Analysis}
\vspace{-5pt}
\subsection{Main Results}
\vspace{-5pt}
Table~\ref{tab:asvspoof_order} compares the proposed models with the
XLSR-AASIST baseline. Cov-ST models spectro-temporal dependencies using only covariance-induced relations. At $K=2$, Cov-ST achieves a competitive EER of 1.15\% on 2021 LA, compared with 1.00\% for the baseline, while reducing the 2021 DF EER from 3.69\% to 2.66\%, a relative reduction of 27.9\%. These results
are obtained with fewer trainable back-end parameters than the baseline.

DuRe-ST jointly models covariance- and attention-induced relations. With
$K=1$, it achieves the best EERs of 0.93\% and 2.04\% on 2021 LA and DF,
respectively. Relative to the baseline, DuRe-ST reduces EER by 7.0\% and 44.7\%,
yielding an average relative reduction of 25.9\% across the two
benchmarks. This improvement requires only 4k additional trainable back-end parameters. Overall, the Cov-ST results demonstrate that covariance-induced relations capture discriminative spoofing cues, while the DuRe-ST results support
their complementarity with attention-induced relations.

\begin{table}[t]
\centering
\caption{ASVspoof benchmark performance and sensitivity to the polynomial order $K$, reported in terms of EER (\%). Params. denotes the number of trainable
back-end parameters in millions. The best results are highlighted in bold.}
\vspace{0.3em}
\label{tab:asvspoof_order}
\resizebox{0.95\linewidth}{!}{
\begin{tabular}{lcccc}
\toprule
\textbf{Model}
& $\boldsymbol{K}$
& \textbf{Params. (M)}
& \textbf{2021 LA}
& \textbf{2021 DF} \\
\midrule
XLSR-AASIST~\cite{tak2022wav2vec}
& -- & 0.447 & 1.00 & 3.69 \\
\midrule
\multirow{4}{*}{Cov-ST (Ours)}
& 0 & 0.387 & 1.05 & 3.15 \\
& 1 & 0.391 & 1.56 & 2.71 \\
& 2 & 0.395 & 1.15 & 2.66 \\
& 3 & 0.400 & 1.08 & 7.79 \\
\midrule
\multirow{4}{*}{DuRe-ST (Ours)}
& 0 & 0.447 & 2.01 & 2.23 \\
& 1 & 0.451 & \textbf{0.93} & \textbf{2.04} \\
& 2 & 0.455 & 1.19 & 2.71 \\
& 3 & 0.459 & 2.17 & 2.41 \\
\bottomrule
\end{tabular}}
\vspace{-12pt}
\end{table}

\begin{figure*}[t]
    \centering
    \includegraphics[width=0.9\textwidth]{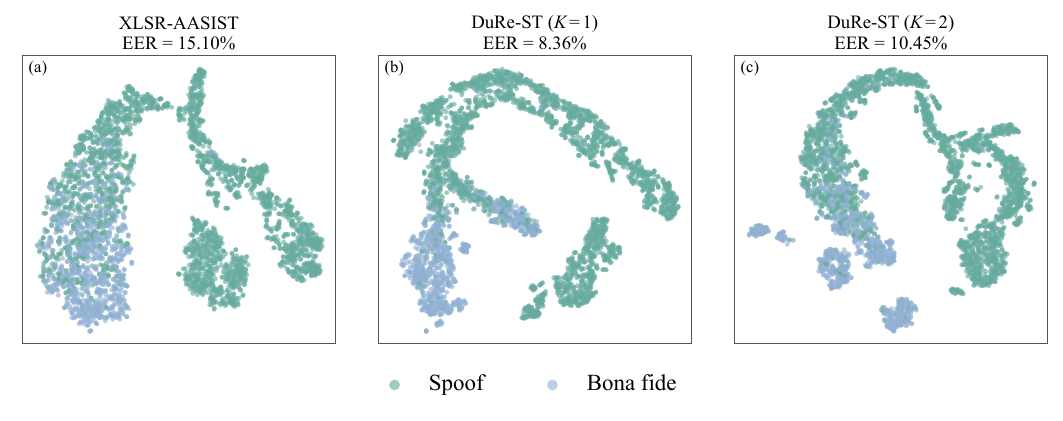}
    \vspace{-12pt}
    \caption{t-SNE projections of utterance-level embeddings on the DFADD evaluation set for (a) XLSR-AASIST, (b) DuRe-ST with $K=1$, and
    (c) DuRe-ST with $K=2$. The corresponding EER is reported above each panel.}
    \label{fig:dfadd-tsne}
    \vspace{-12pt}
\end{figure*}

\begin{table*}[t]
\centering
\caption{Cross-dataset comparison with publicly available SDD models in
terms of EER (\%). $^\dagger$ and $^*$ denote our reproduced results and
results from Speech DF Arena~\cite{leaderboard}, respectively. Among the proposed models, the largest and second-largest EER reductions over the
XLSR-AASIST baseline are shown in boldface and underlined, respectively.}
\vspace{0.3em}
\label{tab:cross_dataset}
\begin{tabular*}{0.9\textwidth}{
@{\extracolsep{\fill}}lccccc@{}}
\toprule
\textbf{Model}
& \textbf{Params. (M)}
& \shortstack{\textbf{In-the-Wild}}
& \shortstack{\textbf{Fake-or-Real}}
& \shortstack{\textbf{ASVspoof 5}}
& \shortstack{\textbf{DFADD}} \\
\midrule
\multicolumn{6}{@{}l}{\textit{Public models}} \\[-2pt]
XLSR-Nes2Net-X~\cite{liu2025nes2net}\arena
& 0.512 & 7.75 & 6.31 & 22.05 & 11.14 \\
XLSR-Mamba~\cite{xiao2025xlsrmamba}
& 1.937 & 6.70 & 6.71 & 14.40\arena & 11.81\reimpl \\
XLSR-Conformer~\cite{rosello2023conformer}\reimpl
& 2.383 & 6.68 & 7.55 & 17.21 & 11.51 \\
XLSR-Conformer-TCM~\cite{truong2024tcm}
& 2.383 & 7.79\arena & 10.68\arena & 18.85\arena & 10.33\reimpl \\
XLSR-SLS~\cite{zhang2024sls}\reimpl
& 23.399 & 9.65 & 7.55 & 18.50 & 9.02 \\
\midrule
\multicolumn{6}{@{}l}{\textit{Baseline}} \\[-2pt]
XLSR-AASIST~\cite{tak2022wav2vec}
& 0.447 & 11.19\arena & 7.46\arena & 16.24\arena & 15.10\reimpl \\
\midrule
\multicolumn{6}{@{}l}{\textit{Ours}} \\[-2pt]
Cov-ST \hspace{0.6em}($K=1$)
& 0.391 & \underline{7.51} & \underline{6.54} & 19.94 & 10.47 \\
Cov-ST \hspace{0.6em}($K=2$)
& 0.395 & 10.20 & 9.50 & 17.88 & 17.36 \\
DuRe-ST ($K=1$)
& 0.451 & 8.56 & 8.22 & \underline{16.10} & \textbf{8.36} \\
DuRe-ST ($K=2$)
& 0.455 & \textbf{6.53} & \textbf{5.48} & \textbf{13.84} & \underline{10.45} \\
\bottomrule
\end{tabular*}
\vspace{-12pt}
\end{table*}

\vspace{-5pt}
\subsection{Sensitivity to the Polynomial Order}
\vspace{-5pt}

The order $K$ determines the highest power of the normalized affinity
matrix included in the polynomial filter. At $K=0$, $\widetilde{\mathbf{A}}^{0}=\mathbf{I}$, so the model retains only an additional transformation of the shared
spectro-temporal representations without using covariance-induced information. Without covariance information, the additional transformation reduces EER
from 3.69\% to 2.23\% on 2021 DF, but more than doubles it from 1.00\%
to 2.01\% on 2021 LA. This substantial degradation on LA shows that the
transformation alone does not yield consistent gains over the baseline. At $K=1$, the filter incorporates covariance-induced relations, and
DuRe-ST achieves its best EERs of 0.93\% and 2.04\% on 2021 LA and DF,
respectively, outperforming both its $K=0$ counterpart and the baseline. Extending the filter
to $K=2$ yields EERs of 1.19\% and 2.71\%, which remain competitive but do
not improve upon $K=1$. This comparison demonstrates that the gains stem from covariance-induced relation modeling itself, rather than merely from introducing an additional transformation of the shared spectro-temporal representations.

The benefit does not increase monotonically with $K$. Cov-ST achieves its best 2021 DF result at $K=2$, but its EER increases sharply to 7.79\% at
$K=3$. DuRe-ST remains comparatively stable on DF, although its 2021 LA EER
also degrades to 2.17\% at $K=3$. While covariance propagation is beneficial,
higher-order filtering may aggregate redundant or weak covariance relations,
thereby diluting discriminative spoofing cues. We therefore retain the
low-order settings $K=1$ and $K=2$ for cross-dataset evaluation.
\vspace{-5pt}
\subsection{Cross-Dataset Comparison with Public SDD Models}
\vspace{-5pt}
Table~\ref{tab:cross_dataset} compares Cov-ST and DuRe-ST with the
XLSR-AASIST baseline and representative public SDD models. With $K=1$,
Cov-ST outperforms the baseline on all datasets except ASVspoof~5 while
using 56k fewer trainable back-end parameters. It is also competitive with
several public models on In-the-Wild, Fake-or-Real, and DFADD. However,
these advantages diminish at $K=2$, which offers only a modest improvement
on In-the-Wild and performs worse than the baseline on the remaining
datasets. At $K=2$, the largest degradation occurs on DFADD, with a 65.8\%
relative increase in EER over $K=1$.
These results indicate that covariance-induced relations provide useful
spoofing cues, but their effectiveness varies across datasets and filter
orders. Higher-order filtering may propagate less informative relations,
weakening the discriminative cues captured by covariance-only modeling.

DuRe-ST exhibits stronger and more consistent performance. With $K=2$, it
outperforms the baseline on all four datasets and achieves the best results
among the compared systems on In-the-Wild, Fake-or-Real, and ASVspoof~5.
Its EERs of 6.53\%, 5.48\%, and 13.84\% improve upon the strongest public
results by 2.2\%, 13.2\%, and 3.9\%, respectively, in relative terms.
On DFADD, DuRe-ST with $K=1$ achieves the lowest EER of 8.36\%, a relative reduction of 44.6\% over the baseline and 7.3\% over the best public result. These gains require only 4--8k additional trainable back-end parameters. With
0.451--0.455M parameters, DuRe-ST remains smaller than all public models
in the comparison, whose back-ends range from 0.512M to 23.399M parameters.

Overall, the Cov-ST results highlight both the discriminative information
captured by covariance-induced relations and the limitations of relying on
them alone. The stronger DuRe-ST results support their complementarity with
attention-induced relations for cross-dataset generalization.

\subsection{Visualization of Embedding Distributions}

Figure~\ref{fig:dfadd-tsne} visualizes the t-SNE projections of
utterance-level embeddings on the DFADD evaluation set. Compared with the
XLSR-AASIST baseline in Fig.~\ref{fig:dfadd-tsne}(a), both DuRe-ST variants exhibit clearer separation between bona fide and spoofed samples. DuRe-ST with $K=1$ in Fig.~\ref{fig:dfadd-tsne}(b) shows the least class overlap,
consistent with its lowest EER of 8.36\%. By comparison, the increased local
overlap observed for $K=2$ in Fig.~\ref{fig:dfadd-tsne}(c) corresponds to its higher EER of 10.45\%. Although t-SNE provides only qualitative evidence, the observed class
separation is consistent with the EER results, suggesting that
covariance-induced relation modeling contributes to a more discriminative
embedding space.

\section{Conclusion}

We have introduced covariance-induced spectro-temporal relation modeling for SDD
and proposed Cov-ST and DuRe-ST. Cov-ST demonstrates that second-order
dependencies provide discriminative spoofing cues, although relying on these
relations alone is sensitive to the polynomial order and evaluation domain.
DuRe-ST achieves an average relative EER reduction of 25.9\% over the
baseline on the ASVspoof benchmarks and outperforms the strongest public
models in the comparison by 2.2--13.2\% in relative EER, with only
0.451--0.455M trainable back-end parameters. These results support the
complementary role of covariance-induced spoofing cues in improving
cross-dataset generalization through joint modeling with attention-induced
relations. Future work will explore input-adaptive order weighting to reduce order sensitivity.

\newpage

\bibliographystyle{IEEEbib}
\bibliography{reference}

\end{document}